\documentclass[letterpaper]{raa}

\usepackage{amssymb}
\usepackage{multirow}

\usepackage{graphicx}
\usepackage{enumerate}
\usepackage{amsmath}
\usepackage{enumitem}

\usepackage{rotating}
\usepackage{pdflscape}

\usepackage{graphicx,times,amssymb,hyperref}             %for PS/EPS graphics inclusion, new
\input{epsf.sty}                        %for PS/EPS graphics inclusion, old
\input{psfig.sty}                       %for PS/EPS graphics inclusion, old
\usepackage[totalwidth=480pt,totalheight=680pt]{geometry}

\usepackage{enumerate}

\def\cm{\,{\rm cm}}

\def\ergscm2 {erg\,s$^{-1}$cm$^{-2}$}

\def\cm2 {cm$^{-2}$}

\def\aap {A\&A}
\def\apj {ApJ}

\begin{document}

%*********************************************************************
\title{Hints of a second explosion (a quark nova) in Cassiopeia A Supernova}

 \volnopage{Vol.0 (200x) No.0, 000--000}      %%preserved for Editor. DOn't remove!
   \setcounter{page}{1}          %%starting page, preserved for Editor. DOn't remove!

   \author{Rachid Ouyed  %  \inst{1} 
   \and Denis Leahy  % \inst{1}
    \and Nico Koning  % \inst{1}
      }

   \institute{Department of Physics and Astronomy, University of Calgary, 
2500 University Drive NW, Calgary, Alberta, T2N 1N4 Canada; {\it rouyed@ucalgary.ca}
   }

   \date{Received~~2014 month day; accepted~~2014~~month day}
%%%%%%%%%%%%%%%%%%

\abstract{We show that the explosive transition of the neutron star (NS) to a quark star (QS) (a Quark Nova) in Cassiopeia A (Cas A) a few days following the SN proper can account for several of the puzzling kinematic and nucleosynthetic features observed.  The observed decoupling between Fe and $^{44}$Ti and the lack of Fe emission within  $^{44}$Ti regions is expected in the QN model owing to the spallation of the inner SN ejecta by the relativistic QN neutrons.  Our model predicts the $^{44}$Ti  to be more prominent to the  NW of the  central compact object  (CCO) than in the SE and little of it  along the NE-SW jets, in agreement with NuStar observations. Other intriguing features of Cas A such as the lack of a pulsar wind nebula (PWN) and the reported  a few percent drop of the CCO temperature over a period of 10 years are also addressed.
}

   \authorrunning{Ouyed  et al.}            %author_head in even pages
   \titlerunning{Hints of a second explosion (a QN) in Cas A SN}  % title_head in odd pages

   \maketitle

%*********************************************************************

\section{Introduction}

Cassiopeia A (Cas A) is a young ($<$ 400 years) near-by ($\sim 3.4$ kpc) supernova (SN)\footnote{Its spectrum shows that it was a type IIb SN  (Krause et al. 2008).} remnant which provides an excellent test-bed for SN explosion models.    Recent observations  by NuStar (Grefenstette et al. 2014)  find that $^{44}$Ti is mostly interior to the reverse shock.  On the other hand,  Hwang \& Laming (2012)  noted that the Fe-rich ejecta is well outside the reverse shock\footnote{Earlier analysis pointed out that Cas A's Fe emission differs from that of the lower-Z species (Hughes et al. 2000).}.
This   spatial disconnect between $^{44}$Ti and Fe  places our understanding of SN nucleosynthesis at odds with current explosion models. 
 
Grefenstette et al. (2014) suggested that this lack of spatial correlation between Fe and $^{44}$Ti might be due to additional Fe-rich ejecta having not yet been heated by the reverse shock and therefore quiet in X-rays. 
   Delaney et al. (2014) however have measured the total amount of un-shocked ejecta to be $\sim 0.39M_{\odot}$  (close
  to predicted values; Hwang \& Laming 2012) using low-frequency radio absorption; a small mass compared to that of the shocked ejecta.  Furthermore, there does not seem to be any  significant amount of Fe  in the un-shocked ejecta
  since it would have been visible to Spitzer (e.g. Delaney et al. 2010). As suggested by Grefenstette et al. (2014), a viable alternative is that most of the Fe is already shocked and visible, and that some mechanism decouples the formation of $^{44}$Ti and Fe, thereby producing the observed uncorrelated spatial map (see  Laming 2014; see also Hwang \& Laming 2003).

 Recent 3-dimensional maps derived from observations (e.g.  DeLaney et al. 2010; Milisavljevic \& Fesen 2013 to cite only a few) find morphological features unlike what one would expect from symmetric jet-driven explosions. For example, the NE-SW streams of ejecta (the so-called Cas A jets with velocities exceeding $+10000$ km s$^{-1}$), have observed opening half-angles of ($\sim$ 40$^{\rm o}$) which is much higher than the $< 10^{\rm o}$ opening half-angle expected in jet-induced explosions (e.g., Khokhlov et al. 1999; Wheeler et al. 2002; Akiyama et al. 2003; Fryer \& Warren 2004; Laming et al. 2006).  The CCO is moving with a transverse velocity kick of 350 km s$^{-1}$ in a direction nearly \textit{perpendicular} to the jet axis.     If  the NE-SW axis is the progenitor's rotation axis one might expect the CCO to be kicked in the direction  roughly aligned with the jet axis\footnote{Strong rotation leads to  the CCO kick along the jet axis.
For slow rotation, such that the accretion shock remains spherical, instabilities grow strongest in the equatorial plane, which should give   a kick orthogonal to the jet axis (Yamasaki \& Foglizzo 2008; Iwakami et al. 2009).} (e.g. Burrows \& Hayes 1996; Fryer \& Warren 2004).
Alternatively, Wongwathanarat,  Janka, \& M\"uller (2013) show that
gravitational forces result in a kick direction toward the slowest moving
clumps of ejecta. In this case, the CCO  kick is not expected to be aligned with the jet axis
(i.e. presumably the direction of the fastest moving ejecta). If 
 there is asymmetry in the matter perpendicular to the jet axis, the kick
would therefore also be perpendicular to the jets (see also Janka et al. 2005).

   Asymmetries were found using Light Echoes (LE) which 
show higher velocities in the NW compared to  other  directions (Rest et al. 2011).
 Hwang \& Laming (2012) measure the bulk ejecta center-of-mass moving in the opposite direction to the CCO kick (i.e. faster
 in the N), similar to the LE.
NuStar, Chandra  and LE observations coupled with Doppler maps and 3D modelling of Cas A depict an ejecta unlike that expected in traditional SN explosion models. The ejecta  shows a complex morphology
  with different geometries (outflows, rings etc...) which remain to be explained.

  We provide an alternative physical scenario which appears to be consistent with the existing observational findings: the Cas A remnant is the result of two explosions.  The SN   proper leaves behind a neutron star (NS) which then explodes a few days later as a Quark Nova (QN) leaving behind a quark star (QS).  \\

The basic picture we present here consists of:

\begin{enumerate}[label=\roman*]

{\it 

\item An asymmetric SN explosion where the CCO kick direction is opposite to the direction of the bulk
of the nucleosynthetic products  (e.g. by the gravitational tug-boat mechanism; Wongwathanarat et al. 2013).

\item  A NS is left behind with its PWN.  We assume a NS born with a moderate  period
of a few millisecond and adopt $P_{\rm NS}=  2$ ms as our fiducial value. Faster periods are
expected mainly when rotation  affects significantly the explosion (e.g. Janka et al. 2005).

\item  We assume that the SN explosion  left behind a relatively massive NS and  take $M_{\rm NS}=2M_{\odot}$ as a fiducial value.
 This is not unreasonable for the $\sim$ 15-25$M_{\odot}$ progenitor in Cas A (e.g. Young et al. 2006).

\item A symmetric QN explosion occurs a few days following the SN. 
The QN has three main effects:
(a) It turns off the PWN; (b) It blows out the end-caps of the elongated PWN to give the NE-SW jets;
(c) It spallates (i.e. destroys) the inner $^{56}$Ni   leaving $^{44}$Ti  as the main
imprint of the original post-SN $^{56}$Ni distribution.

}

\end{enumerate}

The interaction of the neutron-rich relativistic QN ejecta (with Lorentz factor $\Gamma_{\rm QN}\sim 10$) with the preceding SN ejecta in Cas A leads to spallation of the inner $^{56}$Ni and turns it into $^{44}$Ti and other spallation products  (Ouyed et al. 2011).  
 In our model, a time delay ($t_{\rm delay}$) of a few days between the SN and QN explosions
 is optimal to account for the formation and abundance\footnote{For time delays of a few days  the amount of $^{44}$Ti produced by spallation of the inner $^{56}$Ni  is of the order of $10^{-4}M_{\odot}$, similar to 
measured values in Cas A (Iyudin et al. 1994; Vink et al. 2001).} of $^{44}$Ti in Cas A (Ouyed et al. 2011).

Here we provide further evidence that the nucleosynthesis and  kinematics in the Cas A ejecta and the properties and cooling of the CCO find natural explanation in the dual-explosion model. 
 Besides helping resolve the $^{44}$Ti-Fe conundrum, our model as we show in this work also explains the absence of the PWN and 
accounts for the rapid cooling of the CCO.    

This paper is outlined as follows: In \S \ref{sec:qn} we introduce the reader to key features of the QN and focus on the properties of its ejecta and its compact remnant. In \S \ref{sec:CasA} we investigate the impact of a QN occurring a few days following the SN explosion in Cas A. In particular, we show how the observed asymmetries and the cooling behaviour of the CCO in Cas A can be explained in our model. We list some predictions in \S \ref{sec:pre} before we conclude in \S \ref{sec:con}.
 
\section{The Quark Nova}
\label{sec:qn}

The QN is the explosive transition of a NS to a QS (Ouyed et al. 2002; Ker\"anen et al. 2005; Vogt et al. 2004; Ouyed\&Leahy 2009; Niebergal et al. 2010a; Ouyed et al. 2013a).  
A relatively massive NS with birth period of a few milliseconds would  spin-down to increase its core density to 
 quark deconfinement value  on timescales of a few days if its 
 magnetic field is $B_{\rm NS}\sim 10^{14}$ G  (e.g. Table 2 in Staff et al. 2006). 
 For our fiducial values, the NS would experience a QN explosion in $\sim$ 5 days
 at which point the NS would have spun-down to a period of $\sim 4$ ms (see Figure 4 in Staff et al. 2006).
 Alternatively, the  NS could experience a QN event following accretion which would drive the 
 NS  above the critical mass sustainable by neutron matter. 
 However, the 5 day time delay is not easily explained in this picture since SN fall-back is expected to occur much sooner than that. 
 We note that the outward propagation of the quark core (made of up, down and strange quark matter) and the ensuing QN occurs in a matter of seconds following quark deconfinement (see Niebergal et al. 2010a and Ouyed et al. 2013a for a recent review).

~\\

We list below the properties of the QN ejecta and its compact remnant that are relevant to Cas A:
 
\begin{enumerate}[label=\roman*]

\item \textit{The QN ejecta}: The QN ejects the outermost layers of the NS with kinetic energy exceeding $\sim 10^{52}$ erg.  On average $10^{-3}M_{\odot}$ of iron-rich and neutron-rich material is ejected at relativistic speeds (with a typical Lorentz factor of $\Gamma_{\rm QN}\sim 10$).

\item \textit{r-process material}:  The neutron-rich QN ejecta and the seed nuclei present in the expanding NS crust, provide ideal conditions for a robust and successful r-process to occur (Jaikumar et al. 2007; Charignon et al. 2011; Kostka et al. 2014a\&b). Heavy elements with atomic weight $A>130$ are produced with similar abundances.

\item \textit{An aligned rotator}:  The compact remnant (the QS) is born as an aligned rotator (Ouyed et al. 2004) owing to the QS entering a superconducting state (see Ouyed et al. 2006). The magnetic field inside the QS is confined to vortices aligned with the spin axis. No persistent radio pulsations are therefore expected from the compact remnant.    

\item \textit{Fall-back material}: Fall-back QN debris (which amounts to $\sim 10^{-7}M_{\odot}$ of heavy elements) in the close vicinity   of (i.e. a few stellar radii away from) the QS.  
 The fall-back material in the QN is reflective of the ejecta composition; i.e. rich in heavy elements.
The initial composition of the QN ejecta is representative of matter in the outer layers of the
NS which consists of the crust (with mass $\sim 10^{-5}M_{\odot}$ dominated by iron-group elements and neutron-rich  nucleon beyond
iron; Baym et al. 1971) and the neutrons. The mass in the QN ejecta is on average of the order of $10^{-3}M_{\odot}$ (Ker\"anen
et al. 2005; Niebergal et al. 2010) which makes it very neutron-rich. As  the QN ejecta expands, the r-process takes effect  and is completed much before fallback is triggered leading to the formation of heavy elements with atomic weight $A>130$ (Jaikumar et al. 2007; Charignon et al. 2011; Kostka et al. 2014a\&b).  The very neutron-rich QN ejecta means that the ejecta remains neutron-rich even after the r-process phase is completed. 
  Calculations (e.g. Chevalier 1989) and simulations (Kifonidis et al. 2003)  of fallback mass in SNe estimate a fallback mass  of the order of a  few percent of the ejecta mass. Assuming that roughly  1\% of the ejecta falls-back onto the QS we expect about $\sim 0.01\times 10^{-5}M_{\odot} \sim 10^{-7}M_{\odot}$ of the crust material to end up magnetically suspended (the shell) at a few stellar radii from the star (Ouyed et al. 2007a\&b). 
Most of the fallback material ($\sim 0.01\times 10^{-3}M_{\odot} \sim 10^{-5}M_{\odot}$) being neutrons will not be affected by the QS magnetic field and would fall directly onto the QS.

 \item \textit{The QS kick}:    The QN explosion (if asymmetric) can provide a kick to the QS. 
 For  $M_{\rm QN} \sim 10^{-3}M_{\odot}$ relativistic ($\Gamma_{\rm QN}=10$) QN ejecta and a $\sim$ 10\% asymmetry, the QN kick is of the order of   100 km s$^{-1}$. The QN kick, if it occurs, is in addition to the kick acquired by the NS from the SN explosion.

\end{enumerate}

\subsection{Dual-shock QNe}

If the time delay ($t_{\rm delay}$) between SN and QN explosions is too long the SN ejecta will have dissipated such that the QN essentially erupts in isolation. However, when $t_{\rm delay}$ is on the order of days to weeks the QN ejecta interacts and collides
with the preceding SN ejecta creating a dual-shock QN.  For our fiducial values,  it would take the QN ejecta on average a few hours ($\sim v_{\rm SN}t_{\rm delay}/c $) to catch up with the preceding SN ejecta; $v_{\rm SN}$  is the  SN expansion velocity and c the speed of light.

\subsubsection{Spallation efficiency}
\label{sec:efficiency}

Spallation of the inner SN ejecta by the QN neutrons is more prominent for time delays that are of the order of a few days.
Specifically, the number of spallation collisions $n_{\rm coll.}$ depends on the density in the inner SN ejecta
 (i.e.  $^{56}$Ni) when it is hit by the QN neutrons.  Equation (1) in Ouyed et al. (2011) shows that
  $n_{\rm coll.}\propto (v_{\rm SN} t_{\rm delay})^{-2}$ which means that for a given time delay
  ($t_{\rm delay}$) between the two explosions, the inner SN ejecta with the lowest expansion velocity ($v_{\rm SN}$)
  (thus with the highest density) will experience more spallation collisions; i.e. efficient spallation.  
  
   Efficient spallation means
   more $^{56}$Ni destruction and more production of light nuclei (such as H, He and C) than $^{44}$Ti (see Figures 1 \& 2 in Ouyed et al. 2011).  Less efficient spallation on higher velocity ejecta is more likely to produce $^{44}$Ti relative to light elements at the expense of $^{56}$Ni. Thus, in an asymmetric SN explosion, as assumed here,  regions of inner SN ejecta   with higher expansion velocity would produce mainly $^{44}$Ti and less light nuclei and more $^{56}$Ni surviving destruction.

    ~\\
Of relevance to Cas A:

\begin{enumerate}[label=\roman*]

\item \textit{Spallation products}:  $^{44}$Ti is a spallation by-product resulting from the interaction between the relativistic QN neutrons and the preceding SN ejecta. In particular, for time delays of a few days,  we find that $^{44}$Ti is one of the main spallation by-products on $^{56}$Ni and amounts to $\sim 10^{-4}M_{\odot}$ (see Figure 2 in Ouyed et al. 2011):   a typical 10 GeV QN neutron (i.e.
a $\Gamma_{\rm QN}=10$) induces
a multiplicity of $\sim $ 13 which yields a spallation by-product peaking at around  $A\sim (56-13) =$ 43.
  
  High expansion velocity (i.e. low density $^{56}$N regions; the NW ejecta in Cas A) would
   experience less efficient spallation which would produce  $^{44}$Ti relative to light elements 
   (H, He and C) at the expense of $^{56}$Ni. Regions with low expansion velocity (the SE ejecta
   in Cas A) would experience efficient spallation which depletes more $^{56}$Ni to form light
   nuclei and less $^{44}$Ti.  These light nuclei should be   adjacent to $^{44}$Ti in the SE. 
   Overall, more $^{56}$Ni would survive spallation in the NW, with more $^{44}$Ti production, than in the SE.

  \item  \textit{The second shock break-out}:  The QN shock (resulting from the collision
  between the QN and SN ejectae)
   would  break out of the SN ejecta  on timescales of the order of the time delay between the
   two explosions  (i.e. a few days to a few weeks; see \S 3 in Leahy \& Ouyed 2008).  In comparison, the SN shock break-out occurs on timescales of the order of a few hours following core-collapse.  The QN shock will reheat and ionize the SN ejecta
   with unique signatures as discussed later.
   
   \item \textit{QN Energetics}:  The QN injects about $10^{52}$ ergs in kinetic energy. We estimate only about $\sim 10^{50}$ ergs of the energy is used up in spallation\footnote{Each neutron starts out with $\Gamma\sim 10$ ($\sim$ 10 GeV energy) and experiences collisions on heavy nuclei ($^{56}$Ni) to lose energy in stripping off neutrons and protons at $\sim$ 7-8 MeV per nucleon. To  get to $^{44}$Ti, the loss is 12 nucleons or about 100 MeV, while spallating all the way to He, the loss is  50 nucleons per $^{56}$Ni or a few
hundreds of MeV. This is much  less than the   10 GeV which leaves most of the energy in kinetic energy of the spallated neutrons
and protons. Once the kinetic energy per particle drops below  $\sim$ 30 MeV  the spallation stops (see Ouyed et al. 2011) and the energy is dissipated in heat.}.  First, we consider the case that the  bulk of the $10^{52}$ ergs goes
into heating the much more massive SN ejecta. Because the heating occurs early, much of the heat energy is lost
to adiabatic expansion, i.e. converted to kinetic energy of the SN ejecta.   For a few solar-mass   SN ejecta
this would lead to expansion velocities exceeding $\sim 10,000$ km s$^{-1}$, higher than observed. 
Secondly, we include  radiative losses, the main one being the QN shock break-out.
This would heat the photosphere to X-ray ($\sim$ keV) emitting temperatures. Because of the very large size of the photosphere ($> v_{\rm SN}t_{\rm delay}\sim 10^{14}$ cm) when
shock break-out occurs, a significant fraction of the $10^{52}$ ergs is radiated quickly. A blackbody photosphere
would radiate the energy in less than a few seconds but in reality the losses are controlled by photon
diffusion (see Leahy \& Ouyed 2008; Ouyed et al. 2012).  The resulting X-rays and UV photons would be mostly absorbed by the surrounding interstellar medium (ISM) and may be not be visible in optical LE. For example, if half of the energy was radiated in the X-ray shock breakout (with a temperature in the $10^10$ K range) at large radius,  this radiation  could be dissipated in a huge volume with little observable effect.  Assuming Thomson scattering  we estimate the
photon mean-free-path to be of the order of a few kilo-parsecs for ISM densities of  10-100 cm$^{-3}$. 
This translates to heating this volume to temperature of  only $\sim$ 10-100 K.  In addition, the 
ionization of surrounding ISM would be brief and no longer observable. We mention that 
Dwek \& Arendt (2008) interpret the IR light echoes in Cas A as possible shock breakout radiation absorbed by dust and reemitted in the IR.  Further analysis of these echoes might  be useful in  constraining our model.

 \item \textit{The lack of double-hump lightcurve in Cas A}:   For delays of a few days between the SN and the QN explosions,
  the QN energy is channeled into PdV losses and no re-brightening of the SN is expected. The QN is effectively ``buried" in the
  SN explosion.   For time delays of the order of a few weeks,  PdV losses are reduced. In this case, the reheated SN ejecta can radiate at higher levels for longer periods of time. This effectively
 re-brightens the SN ejecta creating a super-luminous SN (Leahy \& Ouyed 2008) with its double-humped light-curve (Ouyed et al. 2009); 
the first hump corresponding to the core-collapse SN proper and the second more prominent hump corresponding to the re-energized SN ejecta (Ouyed et al. 2009).  For  time delays exceeding many weeks  the QN effectively occurs in isolation with no collision between the SN and QN ejecta. Contenders for the double-humped lightcurve are the super-luminous supernova SN2006oz (Ouyed \& Leahy 2013),  SN2009ip and SN2010mc  (Ouyed et al. 2013b).

 \item \textit{The sub-luminous nature of Cas A}:  In our model, the destruction of $^{56}$Ni on timescales 
 shorter than  $^{56}$Ni decay timescale (8.8 days) would  lead to a sub-luminous SN with a weak 77-day Cobalt tail.
 The fact that Cas A was not noticed in the sky in the late 1600s led to the suggestion that 
it may have been a sub-luminous SN; it has been linked to a 6th magnitude star by Flamsteed.
However, an alternative explanation is that the extinction might be high.
   
\end{enumerate}

\subsection{QS Properties}

 As described earlier, the parent NS will experience a QN event and leave behind a QS.
  The resulting QS will have a birth period of $P_{\rm QS, 0}\sim 4$ ms  and a magnetic field $B_{\rm QS, 0}\sim 10^{15}$ G.
   Studies of magnetic field
amplification in quark matter, shows that $10^{15}$ G magnetic fields are readily achievable during the transition from a NS to a QS (Iwazaki 2005). Furthermore, as it cools the QS enters the superconducting Color-Flavor-Locked (CFL) state and becomes crustless\footnote{The CFL being rigorously electrically neutral does not allow hadronic matter (Rajagopal\&Wilczek 2001). We assume that there is no depletion of strange quarks at the surface of the QS (Usov 2004 and references therein).}.  Inside the QS the magnetic field  is confined  to vortices parallel to the rotation axis (Ouyed et al. 2004). As the QS spins-down,  the rotational vortices (including the magnetic field confined in them) are slowly expelled  which leads to continuous magnetic reconnection in the QS equatorial region (see Ouyed et al. 2006). The X-ray luminosity resulting from vortex expulsion and the subsequent decay of the magnetic field via magnetic reconnection is (see Ouyed et al. 2007a\&b) 
 
\begin{equation}
L_{\rm X, v} \sim 2\times 10^{33}\ {\rm erg\ s}^{-1}\ \eta_{\rm X, 0.01} \dot{P}_{\rm QS, -11}^{2}\ ,
\end{equation}

\noindent with $\eta_{\rm X}$ being the X-ray conversion efficiency (in units of 0.01) and the period derivative $\dot{P}_{\rm QS}$ given in units of $10^{-11}$ s s$^{-1}$; the subscript $v$ stands for ``vortex".  The corresponding blackbody (BB) QS temperature is found from $4\pi R_{\rm QS}^2 \sigma T_{\rm QS}^4 \sim L_{\rm X, v}$, 
   
\begin{equation}
\label{eq:TQS}
T_{\rm QS}\sim 0.11\ {\rm keV}\ \eta_{\rm X, 0.01}^{1/4} \left( \frac{\dot{P}_{-11}}{R_{\rm QS, 10}}\right)^{1/2}\ ,
\end{equation}

 \noindent where the QS radius is in units of 10 km and $\sigma$ the Stefan-Boltzmann constant. The derived period ($P_{\rm QS}$), period derivative ($\dot{P}_{\rm QS}$) and magnetic field ($B_{\rm QS}$) of the QS evolve in time as  (Niebergal et al. 2006; Niebergal et al. 2010b)

\begin{eqnarray}
\label{eq:QS-properties}
  P_{\rm QS}   &=&  P_0 \left(1+ \frac{t}{\tau_0}\right)^{1/3} \\\nonumber
  \dot{P}_{\rm QS} &=& \dot{P}_0 \left(1+  \frac{t}{\tau_0}\right)^{-2/3}  \\\nonumber
  B_{\rm QS}  &=&  B_0 \left(1+ \frac{t}{\tau_0}\right)^{-1/6} \ ,
\end{eqnarray}

\noindent with $\tau_0 = 5\times 10^3\ {\rm s}\ P_{\rm 0, 4}^2 B_{\rm 0, 15}^{-2} M_{\rm QS, 2} R_{\rm QS, 10}^{-4}$ and $\dot{P}_0 = P_0/(3\tau_0)$. Here $P_{0, 4}$  and  $B_{0, 15}$ are the  period  and magnetic field of the QS at birth in units of 4 milliseconds and $10^{15}$ G, respectively. In our model $PB^2= P_0B_0^2$  is a constant.

\subsection{QS Shell}

The QS will be surrounded by fall-back debris from the QN explosion.  The fall-back material is representative of the QN ejecta and is thus rich in heavy elements (Jaikumar et al. 2007).   The fate of the QN fall-back material ($\sim 10^{-7}M_{\odot}$;
see \S \ref{sec:qn}) depends on its angular momentum, with two possible outcomes.  The first is when the QN fall-back material is formed with enough angular momentum to move into a Keplerian orbit (Ouyed et al. 2007b), forming a rapidly rotating disk.  For a typical amount of debris of $\sim 10^{-7}M_{\odot}$ this would happen if $P_{QS, 0} < \sim 3 $ ms (eq. 8 in Ouyed et al. 2007a; see \S 2 in Ouyed et al. 2007b for more details).

The second scenario, and the one we believe is the most likely in Cas A, is the formation of a co-rotating shell.  For a QS period of $P_{\rm 0}\sim 4$ ms considered here, there is not enough angular momentum to form a Keplerian disk. Instead a co-rotating shell forms (see Ouyed et al. 2007a) supported by the QS's magnetic field. Specifically, the QN fall-back material is kept in equilibrium at a radius where the forces due to the magnetic pressure gradient and gravity are in balance (eqs. 1\&2 in Ouyed et al. 2007a; see also Ouyed et al. 2007b):

\begin{equation}
R_{\rm Shell}\sim 22.4\ {\rm km}\ \sqrt{\sin \theta_{\rm B}} \frac{B_{0, 15} R_{\rm QS, 10}^3}{M_{\rm Shell, -7}^{1/2} M_{\rm QS, 2}^{1/2}}\ ,
\end{equation}

\noindent where the mass of the co-rotating shell is in units of $10^{-7}M_{\odot}$.  The area of the shell is thus $A_{\rm Shell}= 4\pi R_{\rm Shell}^2 \sin(\theta_{\rm B})$ with $\theta_{\rm B}$ (measured from the QS equator) defining a line neutrality above which the gravity vector is no longer mostly perpendicular to the magnetic field vector.  If sections
of the shell are above this line, they are free to break off and fall into the star's poles along the field lines.  Thus, the geometry is such that there is a thin shell (a few kilometers in thickness) at the equator subtending an angle of $2\theta_{\rm B}$, and empty regions at the poles; in the equation above and hereafter we take $\theta_{\rm B}\sim 60^{\rm o}$.
     
The degenerate solid shell is heated from the inside by surface emission from the QS, originating from the magnetic field annihilation from vortex expulsion as the quark star spins down slowly.  The corresponding BB temperature is found from $A_{\rm Shell} \sigma T_{\rm Shell}^4\sim L_{\rm X, v}$ is then:

\begin{equation}
\label{eq:Tshell}
T_{\rm Shell}\sim 0.08\ {\rm keV}\ \eta_{\rm X, 0.01}^{1/4} \left( \frac{\dot{P}_{\rm QS, -11}}{R_{\rm Shell, 20}}\right)^{1/2}\ ,
\end{equation}

\noindent where the shell inner radius is in units of 20 km.  The surface of the shell is a degenerate solid, composed of a wide range of heavy elements with atomic weight $A>130$, which were produced when the neutron star crust was expelled during the QN event (Jaikumar et al. 2007). Since the heavy elements in the QN fall-back material have equal contribution by abundance,   a hard spectrum is to be expected from the shell.  Specifically, the emission from the shell is a continuum radiation, at a temperature given above  as set by the amount of shell heating.  Besides the  featureless\footnote{The QS is crustless since it is in the CFL phase.} continuum emission from the QS, there is emission from the co-rotating shell.  Thus a two-component emission is a natural outcome of the QN model if the viewing angle is favorable :  The shell surrounds the QS except in the polar regions and is solid and degenerate, thus highly optically thick. 
Only when the system is observed along the poles would one observe the two-component (QS+Shell) spectrum 
 while in systems with the QS viewed in the equatorial plane, the QS surface/spectrum may be  shielded.  The 
two BBs may also overlap which would make  it challenging to differentiate the two.  
We do not expect absorption  from the degenerate solid portion of the shell since  no emission from the underlying QS gets through.

\subsubsection{Shell Atmosphere}
\label{sec:atm}
   
The solid degenerate shell has its own very thin atmosphere on the outer radius away from the star.  The atmosphere corresponds to the non-degenerate portion of the shell and its has a thickness set by the strong gravity of the central QS. The height and the density of the atmosphere are (see Ouyed et al. 2007a\&b and Ouyed et al. 2010):

\begin{eqnarray}
H_{\rm atm.} &\sim& 14.3\ {\rm cm} \frac{T_{\rm Shell, keV} R_{\rm Shell, 20}^2}{\mu_{\rm atm.} M_{\rm QS, 2}} \\\nonumber
     &\sim&  1.18\ {\rm cm} \frac{\eta_{\rm X, 0.01}^{1/4} \dot{P}_{\rm QS, -11}^{1/2} R_{\rm Shell, 20}^{3/2}}{\mu_{\rm atm.} M_{\rm QS, 2}} \\\nonumber
   \rho_{\rm atm.}&\sim& 230 \ {\rm g\ cm}^{-3}\ \mu_{\rm atm.} T_{\rm Shell, keV}^{3/2}\\\nonumber
     &\sim&  5.48\ {\rm g\ cm}^{-3}\ \frac{\eta_{\rm X, 0.01}^{3/8} \dot{P}_{\rm QS, -11}^{3/4}}{R_{\rm Shell, 20}^{3/4}}  \ ,
\end{eqnarray}

\noindent where $\mu_{\rm atm.}$ is the atmosphere's mean-molecular weight.  We made use of eq.(\ref{eq:Tshell}) to express $T_{\rm Shell}$ (given in units of keV) in terms of $\dot{P}$ in equations above and below. The mass of the atmosphere is then $M_{\rm atm.}= A_{\rm Shell}\times (H_{\rm atm.}\rho_{\rm atm.})$, or,
   
\begin{eqnarray}
 M_{\rm atm.} &\sim& 7.1\times 10^{-17}M_{\odot}   \frac{T_{\rm Shell, keV}^{5/2} R_{\rm Shell, 20}^4}{M_{\rm QS, 2}}\\\nonumber
   &\sim&  1.4\times 10^{-19}M_{\odot}   \frac{\eta_{\rm X, 0.01}^{5/8} \dot{P}_{\rm QS, -11}^{5/4} R_{\rm Shell, 20}^{11/4}}{M_{\rm QS, 2}}. 
\end{eqnarray}
  
The opacity of the shell's atmosphere, $\tau_{\rm atm.}= \kappa_{\rm es} H_{\rm atm.} \rho_{\rm atm.}$, is then:

\begin{eqnarray}
\tau_{\rm atm.} & \sim & 661.0  \frac{T_{\rm Shell, keV}^{5/2} R_{\rm Shell, 20}^2}{M_{\rm QS, 2}}\\ \nonumber
       &\sim& 1.1 \frac{\eta_{\rm X, 0.01}^{5/8}\dot{P}_{\rm QS, -11}^{5/4} R_{\rm Shell, 20}^{3/4}}{M_{\rm QS, 2}} \ ,
\end{eqnarray}

\noindent with  $\kappa_{\rm es} \sim 0.2$ cm$^2$ g$^{-1}$  the electron scattering opacity.  The atmosphere becomes transparent ($\tau_{\rm atm.} \sim 1$) when the shell has cooled to a temperature:

\begin{equation}
T_{\rm Shell, 1} \sim 0.07\ {\rm keV} \frac{M_{\rm QS, 2}^{2/5}}{R_{\rm Shell, 20}^{4/5}}\ ,
\end{equation}

\noindent or equivalently when the period derivative is of the order of 

\begin{equation}
\label{eq:Pdot-c}
\dot{P}_{\rm QS, 1} \sim  9.5\times 10^{-12}\ {\rm s\ s}^{-1}\  \frac{\eta_{\rm X, 0.01}^{-1/2} M_{\rm QS, 2}^{4/5}}{R_{\rm Shell, 20}^{3/5}}\ .
\end{equation}   

We recall that $\dot{P}_{\rm Shell, 1} =\dot{P}_{\rm QS, 1}$ since the shell co-rotates with the QS. For $\dot{P} <  \dot{P}_{\rm QS, 1}$ the atmosphere becomes optically  transparent revealing the underlying continuum emission from the solid degenerate part of the shell. As we show later, this has an important consequence to the inferred cooling of the CCO in Cas A (see \S \ref{sec:cooling}).

\section{QN Application to Cas A}
\label{sec:CasA}

In this section, we use the QN model to explain several distinctive features of Cas A.
Figure  \ref{fig:panels} illustrates the different stages in our model starting with the
asymmetric SN in panel ``a".  For Cas A,  the inner SN ejecta  would be moving
 at higher speeds in the NW direction than in the SE (i.e. CCO) direction prior to the QN.

\subsection{The NS PWN}
\label{sec:pwn}

Analysis of NS kicks in isolated pulsars suggests that, at least statistically, a considerable degree of alignment between projected spin axes and proper motions exist (e.g. Wang, Lai \& Han 2006 and Ng \& Romani 2007).   This is hard to reconcile with our model where we assume that the direction of motion of the NS is in the SE direction. I.e. that the PW bubble should be blown in the SE-NW direction.  
If the spin-axis and proper motion are aligned (kick-spin alignment) then our model
would require an equatorial NS wind (as in the stripped wind model; e.g Michel 1971; Weisskopf et al., 2000;
 Spitkovsky 2006 to cite only a few) to explain the direction of the NE-SW bubble.  On the other hand, 
there remains the possibility that the NS proper motion was originally in the direction of the NE-SW and
the QN-kick gave it an additional component which acts to reduce the correlation between spin and proper motion.

   During the days prior to the QN event, the NS would have carved out a bubble in the SN ejecta.  During its acceleration phase,
  the size of the bubble can be estimated to be (Chevalier 1977; Reynolds\&Chevalier 1984),

\begin{equation} R_{\rm NS, bub.}\sim 3.2\times 10^{14}\ {\rm cm} \ \left( \frac{v_{\rm SN, 5000}^3 L_{\rm sd, 48}}{M_{\rm SN, 5}} \right)^{1/5} t_{\rm 5}^{6/5}\ ,
\end{equation}

\noindent where the SN ejecta velocity is given in units of 5000 km s$^{-1}$, the mass of the SN shell in units of 5$M_{\odot}$ and the NS spin-down luminosity in units of $10^{48}$ erg s$^{-1}$ for our fiducial values of $P_{\rm NS}=2$ ms and $B_{\rm NS}=10^{14}$ G. The time is given in units of 5 days at which point the pulsar bubble is switched off by the QN.  
 
\begin{landscape}
 \begin{figure}
  \centering
  \includegraphics[scale=0.35]{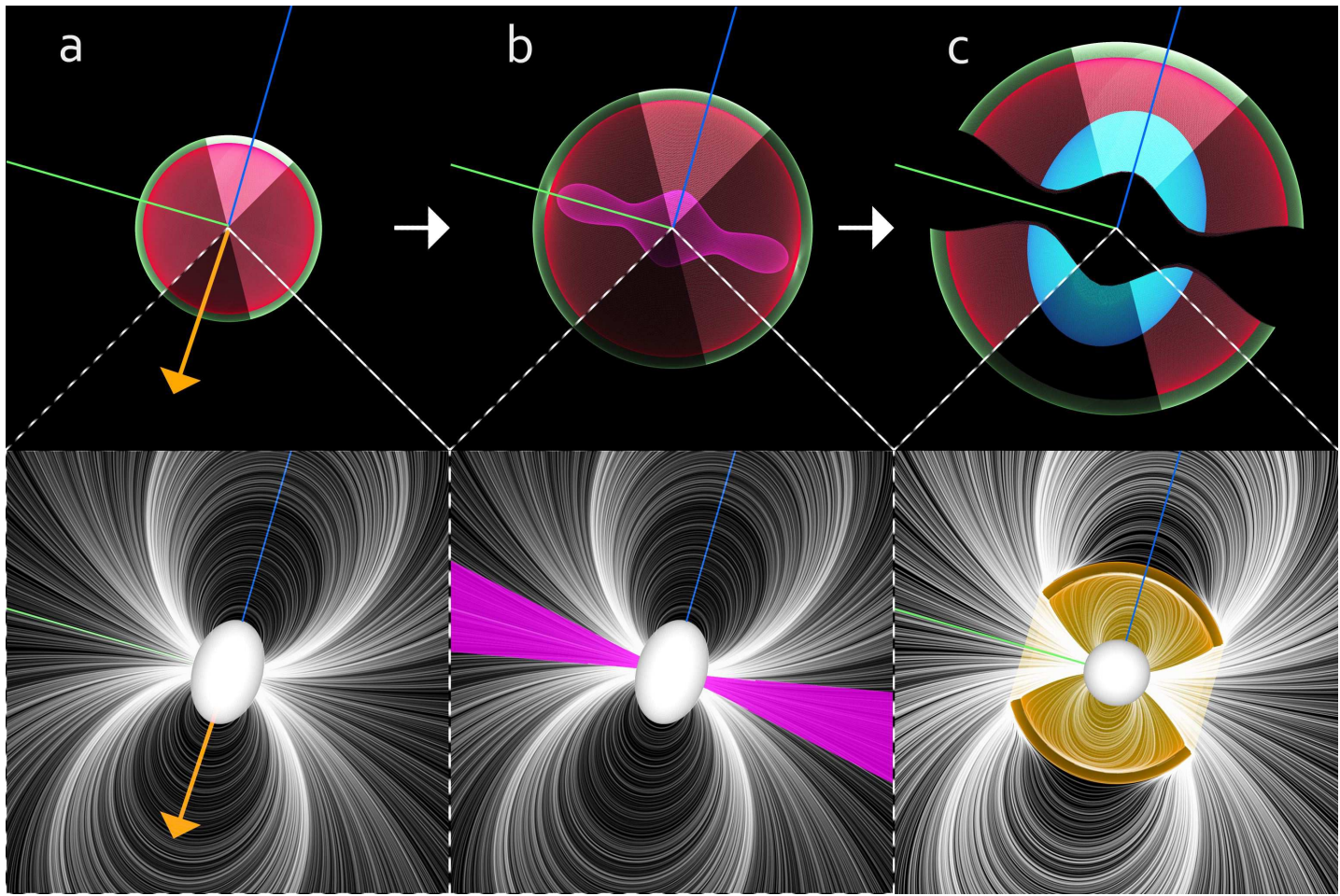}
  \caption{This figure shows a schematic of the evolution of Cas A in the QN model.  In all panels red represents $^{56}$Ni and green Si.  The bottom panels show a close up of the CCO corresponding to the top panel. {\bf Panel (a)} The initial SN explosion is asymmetric with higher velocities toward the NW (represented as a brighter region) and lower velocities toward the SE (dim region).  Due to the asymmetry, the NS is given a kick towards the SE (represented by the orange arrow). {\bf Panel (b)} The rapidly rotating NS blows a pulsar bubble (purple) that is elongated along the rotation axis. {\bf Panel (c)} The second explosion, the QN, shuts off the pulsar wind bubble and clears out a cavity thereby creating the NE-SW jet structure seen in Cas A. $^{44}$Ti is produced in the inner region (blue) by the spallation of $^{56}$Ni. The amount of $^{44}$Ti produced  (and leftover $^{56}$Ni) is lower  in the SE than the NW due to efficient  overall spallation in the SE (see text for details).  The fall-back QN ejecta forms a co-rotating shell around the QS (bottom panel) magnetically supported by the QS B-field.}
\label{fig:panels}
 \end{figure}
\end{landscape}

This can be compared to the size of the spherically expanding SN ejecta, $R_{\rm SN}\sim 2.2\times 10^{14}\ {\rm cm}\ v_{\rm SN, 5000} t_{\rm 5}$.  The $R_{\rm NS, bub.}/R_{\rm SN} \sim 1.5  t^{1/5}$  means that the size of the NS bubble and of the outer SN ejecta evolve in pair.  We expect  a bubble somewhat elongated along the rotation axis (e.g. Bucciantini et al. 2007)  as illustrated  in panel ``b" of Figure  \ref{fig:panels}.

 The PWN would likely be  turned off when the NS converts into a QS.  The crustless QS (being in the CFL phase)  means  a reduced reservoir of   particles to accelerate. Furthermore, 
the conducting co-rotating shell would affect  the magnetic  field line geometry inside the light cylinder (see panel ``c" in Figure \ref{fig:panels}) which we expect would  also  reduce particle injection into a PWN.   This we argue may explain the lack of a  PWN around the CCO in Cas A (Fesen et al.  2006).

\subsection{NE-SW Jets}

 When the QN occurs, only a small fraction, $\zeta_{\rm QN}= \alpha_{\rm NE-SW}^2/4$, of the QN ejecta is launched along the bubble (i.e. the poles of the PWN) carved out by the NS wind.  Here $\alpha_{\rm NE-SW}\sim 40^o$ is the PWN's opening half-angle. Using conservation of angular momentum, $\zeta_{\rm QN} \Gamma_{\rm QN} M_{\rm QN} c\sim  M_{\rm bubble} v_{\rm jet}$, we get
  
\begin{equation}
v_{\rm jet} \sim 3000\ {\rm km\ s}^{-1}\  \frac{\zeta_{QN, 0.1}\Gamma_{\rm QN, 10} M_{\rm QN, -3}}{M_{\rm bubble, 0.1}}\ ,
\end{equation}
 
\noindent where $c$ is the speed of light, $\Gamma_{\rm QN}$ is the QN Lorentz factor in units of 10, and $\zeta_{\rm QN}$ is in units of 0.1. The QN ejecta mass, $M_{\rm QN}$, in units of $10^{-3}M_{\odot}$ and the mass in the bubble $M_{\rm bubble}$ in units of $0.1 M_{\odot}$.  The $M_{\rm bubble}$ can be understood as the amount of bubble material accelerated by the QN ejecta.
   There is a filling factor associated with the  clumpiness of the QN ejecta which means that the QN momentum along the bubble will be imparted to only a fraction of the material in the bubble.
This chunky nature of the QN ejecta (see appendix C in Ouyed \& Leahy 2009) means
 that  not all of the material in the bubble will be cleared out by the QN ejecta. Instead we expect 
 left-over knots from the SN ejecta expanding freely.  Thus along the NE-SW jet, besides the SN ejecta impacted
by the chunky QN ejecta (with some small traces of $^{44}$Ti) there should be some material (chemically and kinematically)  reflective 
 of the SN ejecta.  Detailed numerical simulations of the evolution of the QN ejecta are needed for  a better estimate of the
 filling factor during interaction with the material in the NE-SW bubble.

 Direct mapping of the element abundances  in Cas A jets show
an ejecta  enriched in Si and Mg and relatively poor in Fe (e.g.
Yang et al. 2008). One suggestion is that this composition may be a signature of 
 incomplete explosive Si-burning and that the jet material did not emerge from as deep in the progenitor
 as seems to be the case in other directions of the ejecta (e.g. Khokhlov et
al. 1999; Hughes et al. 2000). In our model,  as the PWN expands, 
 the $^{56}$Ni layer will experience relatively higher compression than  the overlaying layers.
 This means that the  $^{56}$Ni layer will experience more efficient spallation (down to light elements) while shielding the
overlaying layers.  Thus the Si/Mg-rich and Fe-poor composition of the NE-SW jet may be
an indication of the interaction of the PWN with the SN layers prior to the onset of the QN.

Ejecta knots in the NE region of the jet expand at  velocities extending up to 14,000 km s$^{-1}$  (e.g. Fesen et al. 2006).
Milisavljevic \& Fesen (2013)  find no clear kinematic distinction between the NE and SW ejecta in terms of opening half-angle and maximum expansion velocity.  A lower mass  of material in the bubble (say $\sim 0.02M_{\odot}$) yields a velocity of
 15,000 km s$^{-1}$, not very different from observed values. Increasing the QN energy (e.g. the $\Gamma_{\rm QN}$
  and/or $M_{\rm QN}$) would yield higher values as well.

Numerical simulations of MHD jet and neutrino-driven expansion models can produce such jets 
(e.g. Khokhlov et al. 1999; Kotake et al. 2005). However,  in Cas A it seems unlikely that the NE-SW jets  played an important role in the explosion mechanism (Laming et al. 2006).  The QN model provides an alternative explanation
for the direction of the  jets, the kinematics,  and possibly  the unique chemical
composition of the knots in the NE-SW jets as compared to knots in other directions.

\subsection{CCO inferred motion}

 While it is reasonable to assume that the CCO (NS) kick is inherited from the SN explosion,
 below we give an estimate of the CCO (QS) kick following the QN explosion.
   Any asymmetries in the QN explosion could lead to a QS kick.  Assuming a 10\% asymmetry, the kick imparted to the QS can be found using $0.1\times \Gamma_{\rm QN} M_{\rm QN} c\sim M_{\rm QS} v_{\rm QS, kick}$ which yields

\begin{equation}
v_{\rm QS, kick}\sim 150\ {\rm km\ s}^{-1}   \frac{\Gamma_{\rm QN, 10} M_{\rm QN, -3}}{M_{\rm QS, 2}}\ ,
\end{equation}

\noindent for a 2$M_{\odot}$ QS.   The value above is not very different from observed values 
if the CCO kick was only from the QN (e.g. Fesen et al. 2006; see however Delaney\&Satterfield (2013) for an alternative direction).  
 However,  the QN requires that  instabilities have had time to  develop during the conversion
  and  have led to a few percents asymmetries.    It seems therefore that the CCO has most likely inherited the kick from the SN
  as adopted in this work.

\subsection{$^{44}$Ti-$^{56}$Fe separation}

The  time delay  considered here (i.e. a QN going off a few days following the SN) yields $^{44}$Ti trends similar to those observed in Cas A (Ouyed et al. 2011).  The depletion of $^{56}$Ni and the production of lighter elements (Figures 1\&2 in Ouyed et al. 2011) would account for the sub-luminous nature of Cas A and should explain the lack of $^{56}$Ni in the inner ejecta. Thus in our model, $^{44}$Ti and $^{56}$Ni (and thus Fe) would be naturally decoupled in the inner parts of the SN ejecta.

  For a given time delay $t_{\rm delay}\sim 5$ days, lower expansion velocity $v_{\rm SN}$ of the SN ejecta 
 leads to more spallation collisions. For Cas A, a lower expansion velocity in the SE (CCO) direction 
 translates to  an increase in spallation collisions breaking
 out  $^{56}$Ni into light nuclei (H He, and C) thus, resulting in less $^{44}$Ti.
   For the NW, the effects are just the opposite with less overall spallation
   but more $^{44}$Ti. It means  
   less destruction of $^{56}$Ni, and thus more $^{56}$Fe, should be visible in the NW 
   (see panel ``c" of Figure \ref{fig:panels})  and less light nuclei than in the SE.
   
 Interestingly, the asymmetric $^{44}$Ti distribution
predicted by our model  follows what Nustar sees. Looking at Figure 2 in Grefenstette et al (2014), one can see that
most of the $^{44}$Ti is on one side of the CCO, the opposite side to the kick direction.
 Wongwathanarat, Janka, \& M\"uller (2013)  found 
that the majority of the iron-group elements,  in particular $^{56}$Ni,  and including $^{44}$Ti, will be 
predominantly produced and ejected mostly in large clumps in the hemisphere that points away
from the NS kick-velocity vector. However, it remains
to be shown that $^{44}$Ti and $^{56}$Ni can be spatially separated in their model.

In summary, and in our suggested  scenario  then,  the  asymmetrical SN would eject the $^{56}$Ni into specific regions
at different velocities, followed by a QN that reprocesses some of the nucleosynthetic products.   Thus the $^{44}$Ti distribution and intensity should be a reflection not only of
 the regions of $^{56}$Ni spallation but also of the original SN (velocity) asymmetries.  
The light nuclei produced from spallation are indicators of regions of high-density (low  expansion velocity) $^{56}$Ni ejecta since these would experience more spallation collisions when hit by the QN neutrons.

\subsection{CCO Properties}

For an $\sim 330$ years source, and using our fiducial values of $P_0=$4 ms and $B_0=10^{15}$ G for the QS birth period and magnetic field, eq.(\ref{eq:QS-properties}) gives the following CCO (i.e. the QS and the co-rotating shell) parameters:
   
\begin{eqnarray}
   P_{\rm QS}&\sim& 0.28\ {\rm s}\ \alpha_{\rm QS}^{1/3} \beta_{\rm QS} ^{1/3}\\\nonumber
   \dot{P}_{\rm QS}  &\sim&1.0\times 10^{-11}\ {\rm s\ s}^{-1} \ \alpha_{\rm QS}^{1/3} \beta_{\rm QS} ^{1/3} \\\nonumber
   B_{\rm QS}  &\sim& 1.1\times 10^{14}\ {\rm G} \ \alpha_{\rm QS}^{1/3} \beta_{\rm QS} ^{-1/6}\\ \nonumber
    L_{\rm X, v}  &\sim& 2.1\times 10^{33}\ {\rm erg\ s}^{-1}\ \eta_{\rm X, 0.01} \alpha_{\rm QS}^{2/3} \beta_{\rm QS} ^{2/3}\\ \nonumber
    T_{\rm QS} &\sim& 0.11\  {\rm keV}\ \eta_{\rm X, 0.01}^{1/4} \ \alpha_{\rm QS}^{1/6} \beta_{\rm QS} ^{1/6} R_{\rm QS, 10}^{-1/2}\ ,
\end{eqnarray}

\noindent with $\alpha_{\rm QS}= (P_{\rm 0, 4} B_{0, 15}^2)$ and $\beta_{\rm QS} = (M_{\rm QS, 2}^{-1} R_{\rm QS, 10}^4)$. 
f The equation for $T_{\rm QS}$ is arrived at by using Eq. \ref{eq:TQS}  (with $\dot{P}_{\rm QS}$ as given
above) and assumes BB cooling. Similarly, the shell's temperature assumes BB cooling and stems
from  Eq. \ref{eq:Tshell}  (see Ouyed et al. 2007a\&b for details). 
The temperature and the magnetic field on the inside of the shell (given by  $B_{\rm Shell}= B_{\rm QS} (R_{\rm QS}/R_{\rm Shell})^3$) are:

\begin{eqnarray}
   T_{\rm Shell} &\sim& 0.08\  {\rm keV}\ \eta_{\rm X, 0.01}^{1/4}\ \alpha_{\rm QS}^{1/6} \beta_{\rm QS} ^{1/6} R_{\rm Shell, 20}^{-1/2} \\\nonumber
   B_{\rm Shell} &\sim& 1.3 \times 10^{13}\ {\rm G} \ \alpha_{\rm QS}^{1/3} \beta_{\rm QS} ^{-1/6} (R_{\rm QS, 10}/R_{\rm Shell, 20})^{3}.
\end{eqnarray}

The outer surface of the shell (where the atmosphere lies) will be shielded owing to the  high conductivity of the shell; i.e. $B_{\rm atm.} << B_{\rm Shell}$ (see Figure 1).

  The spin-down power is $L_{\rm sd} \sim 5.5\times 10^{35}\ {\rm erg\ s}^{-1}~ B_{0, 15}^{-2}$ 
which goes mainly into relativistic particles.  However, because of the reduced reservoir of particles in our
 model as compared to the NS model (see discussion in \S \ref{sec:pwn})  we would expect an 
X-ray efficiency below the $\sim$ 1\% expected in pulsars (e.g. Chevalier  2000
   and references therein). This implies $L_{\rm X,  sd} < 5.5\times 10^{33}\ {\rm erg\ s}^{-1}$ in our model.
   We think this emission is extended over a large enough region (still small compared to the full extent of Cas A)
 that it is hidden by SN X-rays from Cas A.\\

\subsection{CCO cooling}
\label{sec:cooling}

Spectral fitting of Chandra observations from 2004 and 2005 (Ho\&Heike 2009) shows that the spectrum can be fit with either a H atmosphere of $k_{\rm B}T \simeq$ 244 eV, a He atmosphere with intermediate T, or a C atmosphere with  $k_{\rm B}T \simeq$ 140 eV. Heavier elements (assuming the atmosphere is composed of a single element) are disfavored because they produce line features which should be detectable in the Chandra spectrum. The C atmosphere was favored because it gives a BB radius of 12-15 km, compared to about 5 km for the H or He atmospheres.  Heinke\&Ho (2010) reanalyzed spectra of the Cas A CCO over a 10 year period, assuming their C atmosphere model, and find the temperature dropped by 4\% (with $4.5\sigma$ significance) over that time period.  
 See Elshamouty et al. (2013) and Posselt et al (2013) for a recent analysis of CCO cooling in Cas A.

 Here we argue that the atmosphere is not Carbon, but in fact made of heavy elements. 
The combination of spectra from the QS and shell mimics that of the Carbon atmosphere model. 
We further argue that the sharp drop in temperature may be artificial and is probably induced by the decreasing contribution of the shell's atmosphere to the spectrum. In the case of Cas A, because the spin-axis of the QS is along the NE-SW jets, an observer would see emission from the co-rotating shell's atmosphere.  Using the second expression in eq.(\ref{eq:QS-properties}) (i.e.  $\dot{P}\propto t^{-2/3}$) and eq.(\ref{eq:Pdot-c}) the shell's optical depth becomes unity when the system's age is of the order of:

\begin{equation}
t_{\rm Shell, 1}\sim 374.7\ {\rm years}\ \frac{\eta_{\rm X, 0.01}^{3/4} R_{\rm Shell, 20}^{9/10}  R_{\rm QS, 10}^2(P_{\rm 0, 4} B_{0, 15}^2)^{1/2}}{M_{\rm QS, 2}^{-17/10}} \ .
\end{equation}

This timescale is very close to the estimated age of Cas A ($\sim 330$ years) which means that the shell's atmosphere in Cas A may currently be transiting from an opaque atmosphere to a transparent one. As long as $\tau_{\rm atm.}> 1$, the atmosphere  would appear as one BB.  However, once the shell's atmosphere enters its low optical depth phase ($\tau_{\rm atm.}  < 1$), the emission from the underlying solid degenerate surface would  also contribute. Being from a mixture of the QN heavy elements, the spectrum from the atmosphere is expected to be spectrally harder, as seen in calculations of heavy element atmosphere spectra (e.g.  Ho\&Heinke 2009 for N, O and Fe spectra).

The relevance here is that, as the QS$+$Shell system ages, it naturally cools down ($T_{\rm QS}\propto \dot{P}_{\rm QS}^{1/2} \propto t^{-1/3}$; $T_{\rm Shell}\propto \dot{P}_{\rm QS}^{1/2}\propto t^{-1/3}$) and the optical depth in the shell's atmosphere decreases ($\tau_{\rm atm.}\propto \dot{P}_{\rm QS}^{5/4}\propto t^{-5/6}$ ). This means that the solid (BB) component (we refer to this as the ``surface emission") increases in importance relative to the heavy element atmosphere component.  Effectively, the spectrum is softening because of the decreasing contribution of the spectrally hard heavy-element atmosphere component.  The decreasing opacity of the atmosphere would mimic the temperature ($T$) drop inferred using fixed spectral models. As shown in the Appendix (and related Figure \ref{fig:cooling} and  in Table \ref{table:cooling}),  a series of spectral fits with a fixed spectral model (either BB or e.g. a C-atmosphere model) would give a decreasing temperature, which is an artifact.

\section{Model predictions}
\label{sec:pre}

The model presented in this paper is an example of a dual-Shock QN, where the QN ejecta catches up to and collides with the previously ejected material from the SN.  This model provides strong early-time features including a re-brightening of the light curve (for time delays of a few weeks)\footnote{The corresponding double-humped light-curve was predicted by Ouyed et al.(2009), a few years before it was first reported for SN2006oz (Ouyed \& Leahy 2013), SN2009ip and SN2010mc (Ouyed et al. 2013b). For Cas A the time delay between the QN and the SN explosion according to our model is of the order of a few days. In this case,  a double-humped lightcurve in the optical is not expected  and most of the QN energy is channeled into PdV work (see discussion in \S \ref{sec:efficiency}).  Thus a combination of the short time delay and the destruction of $^{56}$Ni (and thus a reduction
of $^{56}$Co) should lead to a sub-luminous SN.}, and distinct Gravitational Wave signatures (Staff et al. 2012).  Unfortunately some of  these signatures might not be applicable to, or were unobservable for Cas A, and we must  search for other predictions supporting the idea that a QN is responsible for Cas A.\\

 We start with predictions relevant to nucleosynthesis in Cas A:

\begin{itemize}

\item For a time delay of a few days, among the products of spallation on $^{56}$Ni, other than $^{44}$Ti, are Hydrogen, Helium  and Carbon in equal proportion (see Figure 2 in Ouyed et al. 2011).  We predict about $\sim 10^{-4}M_{\odot}$ of light
nuclei  adjacent to  $^{44}$Ti in the inner ejecta of Cas A, particularly in the SE direction.    A wide range of stable isotopes produced by spallation  includes $^{45}$Sc and $^9$F (see Figure 2 in Ouyed 2013).

 \item   The spallation (multi-generation) neutrons add up to a total mass of $\sim 0.1M_{\odot} M_{\rm QN, -3}$ (see Ouyed et al. 2011)\footnote{The spallation Hydrogen  mainly produced in high-density (low  expansion velocity) $^{56}$Ni ejecta amounts
 to $\sim 10^{-2}M_{\rm Ni}$ (Figure 2 in Ouyed et al. 2011).}. These  neutrons will $\beta$-decay after they have traveled a distance of $\sim 10^{13}$ cm from the QS where the magnetic field is of the order of a few Gauss.  The resulting protons will be trapped by the QS magnetic field (with a Larmor radius of the order of a few centimeters) and cool within years (e.g.  Reynolds 1998). 
  In Cas A,   we expect more Hydrogen to be found in the SE (CCO) direction than in the  opposite direction.  On the
 other hand, the reduced spallation towards the NW direction, means fewer spallation  neutrons but these 
would travel a larger distance  before they decay into protons.

\item We predict an overall lack of spallation products, in particular a lack
of light nuclei (e.g. H, He and C)   in knots along the NE-SW jets compared to the rest of the ejecta.
Most of the QN ejecta in the direction of the jets will propagate along the cavity and hit the far end.  
We therefore expect to see some spallation products (in particular $^{44}$Ti) mainly in the end-caps of the jets. 
 This seems to agree  with the Grefenstette et al. (2014)  who found very little $^{44}$Ti in the jets.
 Overall, the chemical composition in the knots along the  NE-SW jets should be reflective 
 of nucleosynthesis in the SN ejecta prior to the QN explosion.

\item Emission lines from the heavy-elements-rich atmosphere (against the BB of the underlying solid degenerate part of the shell) should be detected if an SGR-like burst occur in the CCO in Cas A.  In Koning et al. (2013) we have already argued for a QN (i.e. an atomic) origin of the $\sim 13$ keV line in several AXPs.

\end{itemize}

We now list other more general predictions: 

\begin{itemize}

\item    The QN break-out shock:  Besides the SN shock break-out which would have occurred a few hours
 after core-collapse, there is also the QN shock.
 The time it takes the QN shock to cross the SN ejecta is of the
order of a few days in Cas A, roughly about the same time the SN light curve (LC) would peak (if powered
 by $^{56}$Ni  decay).  Using LE   one might detect (if thin enough dust filaments exist around Cas A; Rest et al. 2011) a change in spectral features as the QN shock goes through the SN ejecta, reheats it and ionizes it.

\item  Magnetar-like behavior is expected from the CCO according to our model\footnote{The possibility of the CCO in Cas A being a magnetar or a QS have been discussed in the literature (e.g. Pavlov\&Luna 2009).  In the QN model, the QS is also a magnetar owing to the $\sim 10^{15}$ G field expected during the transition from hadronic to quark matter (Iwazaki 2005).}. 
  
The magnetic energy due to field decay (following vortex expulsion) is released continuously
over a long timescale (thousands of years) and gives the steady X-ray luminosity of SGRs and AXPs (see Ouyed et al. 2010 and references therein).  Bursting in the QN model is due to chunks of the shell falling onto the QS. After a certain time (of the order of hundreds of years), the magnetic field decays substantially and the entire shell moves in closer to the QS, causing larger sections to be shifted above the line of neutrality (defined by $\theta_{\rm B}$), thus falling and triggering a burst.  

 \item In our model, there is emission from the QS and the co-rotating shell.   Thus a careful
analysis of the spectrum in Cas A should reveal a   two-component  spectra.
  In Cas A, because the QS is viewed in the equatorial plane, we only see the shell (which
is highly optically thick) and  not the QS surface.Thus
 a two-component spectrum may not necessarily be observed.
 
\end{itemize}   

\section{Discussion and conclusion}
\label{sec:con}

Several peculiar features of Cas A have pushed SN explosion models to their limit.  The NuStar observations of $^{44}$Ti may have been the straw that broke the camel's back.  By appealing to a dual-explosion model, we have shown that several issues pertaining to Cas A can be neatly resolved.  

\begin{enumerate}[label=\roman*]

\item The QN ejecta hits the inner edge of the SN remnant creating $^{44}$Ti through the spallation of $^{56}$Ni. This explains why $^{44}$Ti is seen in the inner regions of Cas A whereas Fe is in the outer regions.  We argue that $^{44}$Ti is an indicator 
 of the $^{56}$Ni (and thus of Fe) distribution induced by the SN  explosion. The amount of $^{44}$Ti  produced would vary
 from  NW to  SE   due to differences in expansion velocities (and thus density) of the $^{56}$Ni ejecta.  
 Higher velocity ejecta (the NW) is more likely to produce $^{44}$Ti relative to light elements at the expense of $^{56}$Ni.
 In the SE, more light nuclei and less $^{44}$Ti will be produced and more $^{56}$Ni depletion (see panel ``c" in Figure \ref{fig:panels}).

\item The NE-SW jet is formed by a pulsar bubble from the NS, initiated after the first explosion, the SN proper.  The second explosion, the QN, occurs days later and shuts off the pulsar bubble.  The ejecta also clears out the NE-SW region creating the elongated, wide opening angle feature we observe today.  

\item The peculiar cooling and detection of an atmosphere around the Cas A CCO can be explained with the co-rotating shell present around the QS in the QN model.

\item  In the QN model, the QS is bare and crustless since it is in the CFL phase which is rigorously electrically neutral (Rajagopal \& Wilczek 2001). The lack of  particles to accelerate and the different geometry of the magnetosphere (due to the co-rotating shell)   in our model as compared to a standard NS model may explain the lack of a synchrotron nebula in Cas A (Hwang et al. 2004). This not unrealistic  suggestion requires detailed studies of the electromagnetic structure of crustless quark stars for confirmation.   On the other hand, the fact that the QS is an aligned rotator may naturally explain the non-detection of radio pulsations from the Cas A CCO (McLaughlin et al. 2001; Mereghetti, Tiengo, \& Israel 2002).

 \item  The sub-luminous nature of Cas A (if confirmed) may find an explanation in our model as a consequence
  of the  destruction of $^{56}$Ni on timescales 
 shorter than  $^{56}$Ni decay timescale. We recall that for $t_{\rm delay}$ of a few days the QN energy 
 is buried in the SN ejecta as  PdV work.

\end{enumerate}

These points provide  evidence for the QN model of Cas A.  However, such a novel and unfamiliar idea may require the observation of several predictions presented in section \ref{sec:pre} before it is given serious consideration.

Confirming that Cas A has experienced a QN event will have important  implications to astrophysics and to physics, in particular to Quantum-Chromo-Dynamics. It will confirm the existence of quark stars and demonstrate that superconducting quark matter is the most stable state of matter in the Universe.  In addition it would support  the idea  that quark matter   is capable of generating a magnetar-strength B-field (Iwazaki 2005).

The QN is unique in its ability to peek at quark matter and its properties, and the proximity of Cas A provides an excellent laboratory for this study.  For example, the time delay between the SN and the QN can give us estimates of the quark deconfinement density (Staff et al. 2006), and inferring the energy released by the QN in Cas A will provide vital clues on whether the transition from hadronic matter to quark matter is a first or a second order transition (Niebergal et al. 2010a; Ouyed et al. 2013a).

   The possible implications to nuclear physics and nuclear astrophysics are worth mentioning. The detection of 
   QN r-process elements  (Jaikumar et al. 2007) in Cas A  would  mean that this new r-process site can no longer be ignored.  As a viable and a robust r-process site, a QN can produce on average $10^{-3}M_{\odot}$ of heavy elements per explosion (Ker\"anen et al. 2005; Jaikumar et al. 2007;  Charignon et al. 2011; Kostka et al. 2014a\&b). This is significant even if the QN rate is as low as 1 per a few hundred SNe.

 An  asymmetric SN  followed by a symmetric QN that reprocesses  the SN elemental composition seems to 
  provide an explanation for some intriguing features in Cas A, in particular the $^{44}$Ti-$^{56}$Ni spatial separation.    
  The time delay of a few days between the two explosions is key  which 
  effectively constrains the properties of the NS that experienced the QN.   Another approach is to consider
  a symmetric SN followed by an asymmetric QN. However, we find that
  the scenario we adopted in this paper may be more natural and simpler, although the second scenario cannot be  ruled out
  at this stage;  see discussion in \S \ref{sec:pwn}.  The $^{44}$Ti  is an imprint of the original SN (velocity) asymmetries. Its intensity
    hints at  the density (i.e.  expansion velocity) distribution of the $^{56}$Ni ejecta just before it is hit by the QN.
    
    Our model relies heavily on the feasibility of an explosive transition of
    a massive NS to a QS which seems possible based on one-dimensional simulations (Niebergal et al. 2010a). However, 
    more detailed multi-dimensional numerical simulations  are required to prove or disprove this result (Ouyed et al. 2013a). 
    Furthermore,   a full treatment of the interaction between the relativistic QN ejecta and the non-relativistic $^{56}$Ni target 
    would require  detailed hydrodynamical simulations which include nucleosynthesis calculations. Nevertheless, 
    we hope  that the scenario presented here and our findings  shows that  a QN  in  Cas A is a real possibility and as such
    it warrants further studies.

\begin{acknowledgements}   

We thank J. M. Laming, U. Hwang  and D. Welch for comments which helped improve this paper.
We are grateful for useful comments from an anonymous referee.
This research is supported by  operating grants from the National Science and Engineering Research Council of Canada (NSERC). N.K. would like to acknowledge support from the Killam Trusts.

\end{acknowledgements}

%---------------------------------------------------------------

\begin{table}[h!]
\centering
\caption{Sample fits to the 2000  and 2009  CCO simulated spectra based on Heinke\&Ho (2009) published spectral parameters. The model includes interstellar absorption with $N_{\rm H}=1.74\times 10^{21}$ cm$^{-2}$. }
\begin{tabular}{|c|c|c|c|c|c|c|c|}\hline
ÊCase & $T_{\rm surf.}$ (keV) Ê& $T_{\rm atm.}$ (keV) & $\chi^2$/dof (2000)  & $\chi^2$/dof (2009) &Ê$\tau_{\rm atm.}$ (2000)  &Ê$\tau_{\rm atm.}$ (2009) & percent drop in $\tau_{\rm atm.}$ \\\hline
case 1 &ÊÊÊ0.4 ÊÊÊÊ&ÊÊÊ0.5 ÊÊÊÊÊ&ÊÊÊÊ44.5/52 Ê&Ê37.6/52 ÊÊ&ÊÊ0.444 &0.237 &Ê46\%  \\
case 2 &ÊÊÊ0.35 ÊÊ&ÊÊÊÊÊ0.45 ÊÊÊÊ&ÊÊÊ45.7/52 Ê&Ê39.8/52 ÊÊ&ÊÊ0.202  &0.142 Ê&30\% \\
case 3 &ÊÊÊ0.3 ÊÊÊ&ÊÊÊÊ0.6 ÊÊÊ&ÊÊÊÊÊÊ60.1/52 &ÊÊ55.3/52 ÊÊ&ÊÊ0.0863  & 0.0660  & 24\% \\\hline 
\end{tabular}
\label{table:cooling}
\end{table}
~\\

\begin{appendix}
\section{The temperature drop in our model}

{\it Our hypothesis}:  We are observing a heavy element atmosphere on top of a solid surface (the degenerate part of
the shell), and the atmosphere is only partly optically thick ($\tau_{\rm atm.} \sim 0.5$).
The solid surface emits a BB of lower $T$, and the atmosphere emits a spectrum which is harder, so a variation in $\tau_{\rm atm.}$ (in this case, a decrease) can mimic a decrease
in $T$. In this scenario the observed $T$ decrease is caused by seeing a larger proportion of the flux from the surface as time increases (and $\tau_{\rm atm.}$ decreases).

{\it Our proposed model}: We assume a plane parallel uniform atmosphere.  We 
write the atmosphere spectrum as $A(E)$ and the surface spectrum as $S(E)$. 
Then the total observed spectrum is  $O(E)=(1-\exp(-\tau))\times A(E)+\exp(-\tau)S(E)$.

{\it Analysis method}:  We use XSPEC (command ``fakeit", with the  Chandra ACIS response matrices) to create a simulated CCO spectrum, with counting statistics corresponding to the C atmosphere fit in 2000 (with $T=2.12\times 10^6$ K) and a second simulated spectrum corresponding to the C atmosphere fit in 2009 (with $T=2.04\times 10^6$ K).
The simulated spectra used BB spectra with temperatures (based on Ho\&Heinke (2009)) that mimic the C atmosphere spectra with the T from Heinke\&Ho (2010).  We find that the 2000 C atmosphere spectrum is well fit with a BB with  $T=0.46$ keV
and the 2009 C atmosphere spectrum is well fit with a BB with $T=0.44$ keV.
We then use these simulated spectra as inputs to XSPEC to test our variable $\tau_{\rm atm.}$ model as described above.

\begin{figure}
\centering
\includegraphics[scale=0.18]{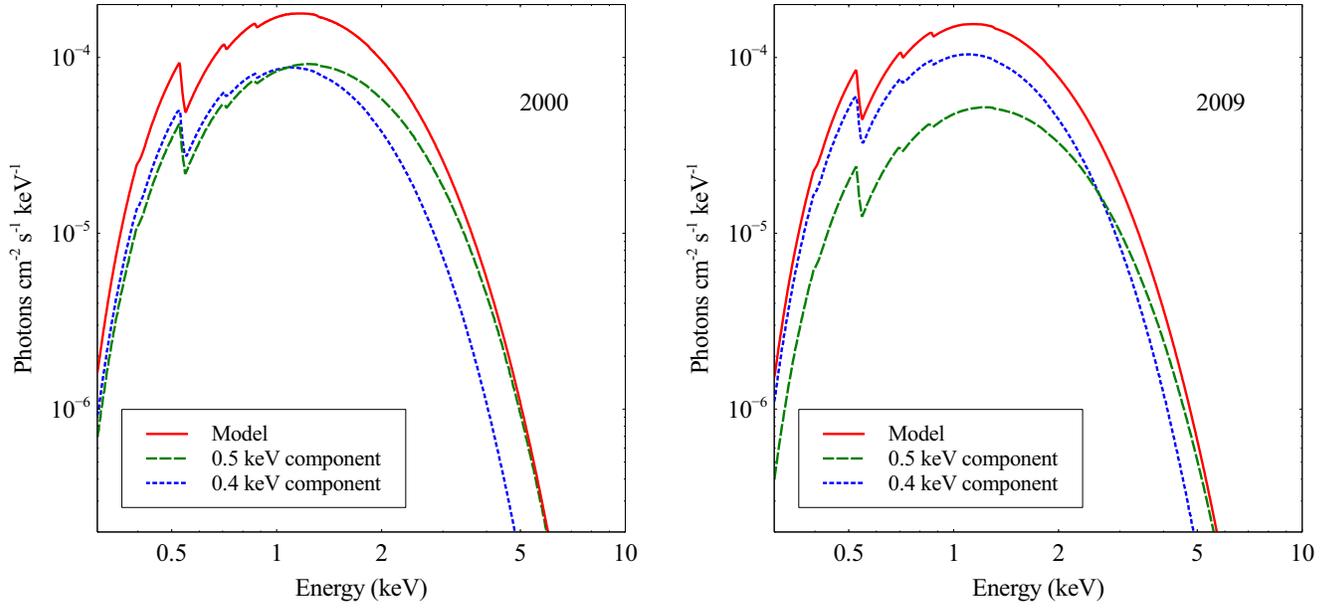}
\caption{Our model fit to the 2000 spectrum (left panel) and the 2009 spectrum (right panel) using a surface (soft BB; $T_{\rm surf.} =0.4$ keV) plus atmosphere (hard BB;  $T_{\rm atm.} =0.5$ keV) with variable optical depth $\tau_{\rm atm.}$. The solid line is the total model spectrum which provides a statistically acceptable fit to the simulated data (see Table in text). The model features are caused by ISM absorption lines. }
\label{fig:cooling}
\end{figure}

{\it Results}:  We confirm our hypothesis as shown in  Figure \ref{fig:cooling} and Table  \ref{table:cooling}. In Figure \ref{fig:cooling},  as an  example, we show the case where  we chose a BB temperature of 0.5 keV to simulate the atmosphere ($T_{\rm atm.}=0.5$ keV) and a BB of 0.4 keV for the BB surface emission ($T_{\rm surf.}=0.4$ keV). In this case the 2000 spectrum could be fit with good statistics with an optical depth of $\sim 0.44$ and the 2009 spectrum with an optical depth of $\sim 0.24$ (see ``case 1" in Table \ref{table:cooling}). This confirms that the apparent temperature drop can be reproduced by a reduction in optical depth  in the 9 year period.  In Table \ref{table:cooling}, we show other cases with different chosen $T_{\rm atm.}$ and $T_{\rm surf.}$.  In our model, since
 $\tau_{\rm atm.}\propto t^{-5/6}$, the drop in the atmosphere's opacity over a decade is of the order of a few percents;
  $\tau_{\rm atm.}(t+\Delta t)/\tau_{\rm atm.}(t)= (5/6) (\Delta t/t) \sim (5/6) (10/330)\sim 0.02$. The discrepancy could be due to some chemical phase change in the atmosphere, such as condensation of the heaviest elements.

\end{appendix}

%%FIGURES

\end{document}